\documentclass[12pt]{iopart}
\usepackage{iopams}  
\usepackage{graphicx}
\usepackage{titlesec}
\usepackage{dcolumn}
\usepackage{appendix}
\usepackage{caption}
\usepackage{geometry}
\begin{document}

\title{Direct evidence of the gradient drift instability being the origin of a rotating spoke in a crossed field plasma}

\author{Liang Xu, Denis Eremin, and Ralf Peter Brinkmann}

\address{Institute for Theoretical Electrical Engineering, Ruhr-University Bochum, D-44780 Bochum, Germany}
\begin{abstract}
A plasma rotating spoke in a crossed field discharge is studied using 2D radial-azimuthal fully kinetic Particle-In-Cell Monte Carlo Collision (PIC/MCC) simulations. The kinetic model reveals the whole perturbation spectrum of the gradient drift instability in the linear stage: Simon-Hoh, lower-hybrid and ion sound modes, providing direct evidence of the spoke of the gradient drift instability nature. The two-fluid dispersion relation of the gradient drift instability was utilized to analyze the linear development of instabilities in the simulations. The charge separation effect was incorporated in the fluid linear theory and a super-resolution signal processing method (multiple signal classification) was applied to obtain the numerical frequency spectrum. The simulated spectrum and growth rate show excellent agreement with the theoretical dispersion relation (real frequency and imaginary frequency) over investigated cases. The most linearly unstable mode was found to be the lower hybrid instability and the mode transition into the m=1 macroscopic rotating structure after saturation of the linear phase is accompanied by an inverse energy cascade. In the nonlinear stage, the pronounced spoke phenomena can occur when the heating of $\mathbf{E_{\theta}\times B}$ electron flow channeled in the spoke front passage suffices to provide the enhanced ionization.

\end{abstract}

\newpagestyle{main}{            
    \setfoot{}{}{\thepage} 
}
\pagestyle{main}    

%
%
%
%

\section{Introduction}
Rotating low frequency, device scale structures referred to as spokes are frequently observed in partially magnetized plasmas, such as Hall thrusters \cite{Janes1966,parker2010,goebel2008}, magnetron discharges \cite{Anders2017,Hecimovic_2015,Brenning_2013} and plasma columns \cite{rebont2011,jaeger2009}. Such plasmas feature crossed external magnetic and electric fields, in which a moderate magnetic field is chosen in such a way that electrons are magnetized but ions are not. Modeling of these discharges is very challenging, because the electron transport across the magnetic field is typically anomalous and the electron mobility derived from the collisional/classical theory does not apply. The rotating spoke is of particular interest since it may play an important role in the anomalous cross-field electron transport \cite{Boeuf2014,Ellison2012}. The spoke rotates as a rigid body with frequency on the order of 10–100 kHz and with the velocity much less than the electron $\mathbf{E} \times \mathbf{B}$ drift. The rotating direction can be either along or opposite to the $\mathbf{E} \times \mathbf{B}$ drift \cite{Hecimovic_2016,Ito2008}.  After the earliest spoke study by Janes and Lowder \cite{Janes1966} and following decades-long investigations, fundamental aspects of the spoke are still poorly understood and operations of Hall plasma devices might be far from predictable and optimal \cite{hagelaar2007modelling,abolmasov2012physics,hara2019overview,Kaganovich2020}. One of the unclear core questions is the driving mechanism behind the formation of the rotating spoke.

So far, there has been no generally accepted mechanism to explain the formation of the rotating spoke in partially magnetized plasmas. Previous researches showed that spokes could be induced through a number of plasma mechanisms. Spokes have been associated with ionization waves and the concept of critical ionization velocity (CIV) proposed by Alfv\'{e}n \cite{Alfven1954} since the early study \cite{Janes1966, Anders2012ionization, Brenning2012,Anders2012apl, piel1980influence, matyash2019}. CIV is the velocity of neutral gas falling through a magnetized plasma reaching the threshold where the kinetic energy of the gas molecules is equal to the energy of ionization. In the context of a rotating spoke, that means the potential drop at the spoke front can provide sufficient energy for electrons to ionize neutrals, form the ionization front and excite an ionization wave.

The spoke may be induced as a result of the modified Simon-Hoh Instability (MSHI) that develops in inhomogeneous partially magnetized plasmas in which the electric field is aligned with a plasma density gradient $\mathbf{E} \cdot \triangledown n>0$ \cite{Simon1963,hoh1963,powis2018,boeuf2019}. The Simon-Hoh instability (SHI) is excited due to different drift velocities of ions and electrons in crossed electric and magnetic fields when the finite resistivity is included. In partially magnetized plasmas, where ions are essentially unmagnetized, the velocity disparity between electrons and ions arises in the azimuthal $\mathbf{E} \times \mathbf{B}$ direction because of the finite ion Larmor radius effect. Therefore, MSHI is also called collisionless Simon-Hoh instability (CSHI) in terms of partially magnetized plasmas. The perturbed azimuthal electric field ($\mathbf{E_{\theta}}$) resulting from CSHI and the enhancement of the density perturbation by the $\mathbf{E_{\theta}} \times \mathbf{B}$ velocity occur in the same manner as in the SHI. It was also claimed that the spoke is the consequence of the linear superposition of several eigenmodes of the gradient drift instability \cite{marusov2019} and the wave coupling of the ion sound wave and the electron Bernstein wave \cite{luo2018spoke}.
In addition, the spoke may result from a nonlinear inverse cascade of short wavelength modes excited by the electron drift instability \cite{sengupta2020}.

The gradient drift instability driven self-organization, structures and anomalous transport in partially magnetized plasmas have been recently demonstrated in two-dimensional fluid simulations \cite{koshkarov2019self,smolyakov2016}. These works verified that the large scale structures, shear zonal flows and vortices, are associated with the inverse cascade of the energy flow from short wavelength
modes. There were also attempts to identify the spoke nature by comparing the gradient drift instability theory with experiments performed in magnetron \cite{Ito2015,Marcovati2020} and hybrid model (fluid electrons and kinetic ions) \cite{kawashima2018numerical}. Further, 2D fully kinetic PIC/MCC simulations with Cartesian grid and artificial ionization were also performed to study micro instabilities and macroscopic structures in the near-anode region of a partially magnetized plasma under conditions typical of Hall thrusters or magnetron discharges \cite{boeuf2019}. In that work, a magnetic field threshold indicating the transition point between the positive sheath and the negative sheath was proposed, and the micro instability development was investigated for different magnetic fields with respect to the threshold value. 

This study contributes to the fundamental understanding of the driving mechanism of the rotating spoke. The work attempts to investigate the gradient drift instabilities and how the small scale modes transit to the rotating spoke by means of a 2D radial-azimuthal fully kinetic PIC/MCC model with self-consistent ionization. Our simulations exhibit a pronounced rotating spoke and a clear linear-nonlinear transition. This enables the precise estimation of instability linear features, e.g., the growth rate and the perturbation spectrum. The comparison between the linear features and theoretical predictions elucidates the mechanism driving the spoke formation. In Sec. 2, the linear theory of the gradient drift instability is briefly reviewed and extended to include the space charge effect. In Sec. 3, we introduce the description of the 2d3v radial-azimuthal PIC/MCC model. In Sec. 4, the observation of the rotating spoke is presented. In Sec. 5, the spectrum and the growth rate are derived and compared to the linear theory. In Sec. 6, the nonlinear saturation and the electron heating will be addressed and the summary is given in Sec. 7.

\section{Overview and extension of the gradient drift instability theory}

A comprehensive overview of the instability theory in inhomogeneous partially magnetized plasmas with slab geometry has been recently given by A. Smolyakov et. al. \cite{smolyakov2016}. This work considers effects of density and potential gradients, electron gyro-viscosity, electron inertia and collisions. The theory applies to $\mathbf{E} \times \mathbf{B}$ plasma configuration with crossed electric and magnetic fields: $\mathbf{E} = E\mathbf{x}$; $\mathbf{B} = B\mathbf{z}$, so that electron drift in the azimuthal direction with velocity $\mathbf{v_{E}} = -v_{E}\mathbf{y} = -E/B \mathbf{y}$, where $\mathbf{x}$, $\mathbf{y}$ and $\mathbf{z}$ are Cartesian unit vectors. The density gradient is in the $\mathbf{x}$ direction, resulting in the electron diamagnetic drift with velocity $\mathbf{v_d}=-v_d\mathbf{y}=-T_e/eBL_n$ where $T_e$ is the elctron temperature, $L_n=n_0/\triangledown n_0$ is the density gradient length and $e$ is the elementary charge. Ions are unmagnetized and can be accelerated in the axial electric field to form the ion beam in the $\mathbf{x}$ direction.

We consider here a purely azimuthal instability, neglect components in $\mathbf{z}$ and $\mathbf{x}$ directions and ions are assumed motionless. With assumptions of quasi-neutrality and uniform $\mathbf{B}$, the dispersion relation writes [formula (31) of Ref. 31]:

\begin{equation}
   \frac{\omega_{d}+k^2_{\theta}\rho^2_{e}(\omega-\omega_{E}+i\nu_{en})}{\omega-\omega_{E}+k^2_{\theta}\rho^2_{e}(\omega-\omega_{E}+i\nu_{en})}=\frac{k^2_{\theta}c^2_s}{\omega^2},  
\end{equation}

\noindent where $k_{\theta}$ is the angular wave number in the azimuthal ($\mathbf{y}$) direction, $\omega$ is the angular frequency, $\rho_{e}=(T_e/m_e)^{1/2}/\omega_{ce}$ is the electron Larmor radius, $m_e$ is the electron mass, $\omega_{ce}$ is the electron cyclotron frequency, $c_s=(T_e/m_i)^{1/2}$ is the ion sound speed, $m_i$ is the ion mass, $\omega_{
d}=k_{\theta}v_{d}$, $\omega_E=k_{\theta}v_E$ and $\nu_{en}$ is the electron-neutral collision frequency. Here, $k_{\theta}$ is in the unit of ${\rm rad/m}$ and $\omega$ in ${\rm rad/s}$. According to the theory, different modes can develop depending on the instability wavelength: the ion sound instability when $k_{\theta}\rho_{e}\gg 1$; the collisionless Simon-Hoh instability when $k_{\theta}\rho_{e}\ll 1$ given $\mathbf{E}\cdot\triangledown n_0>0$; and in-between the lower-hybrid instability. The wave destabilization sources can be the density gradient, the electron $\mathbf{E} \times \mathbf{B}$ drift and collisions. 

\begin{figure}
\center
\includegraphics[clip,width=0.4\linewidth]{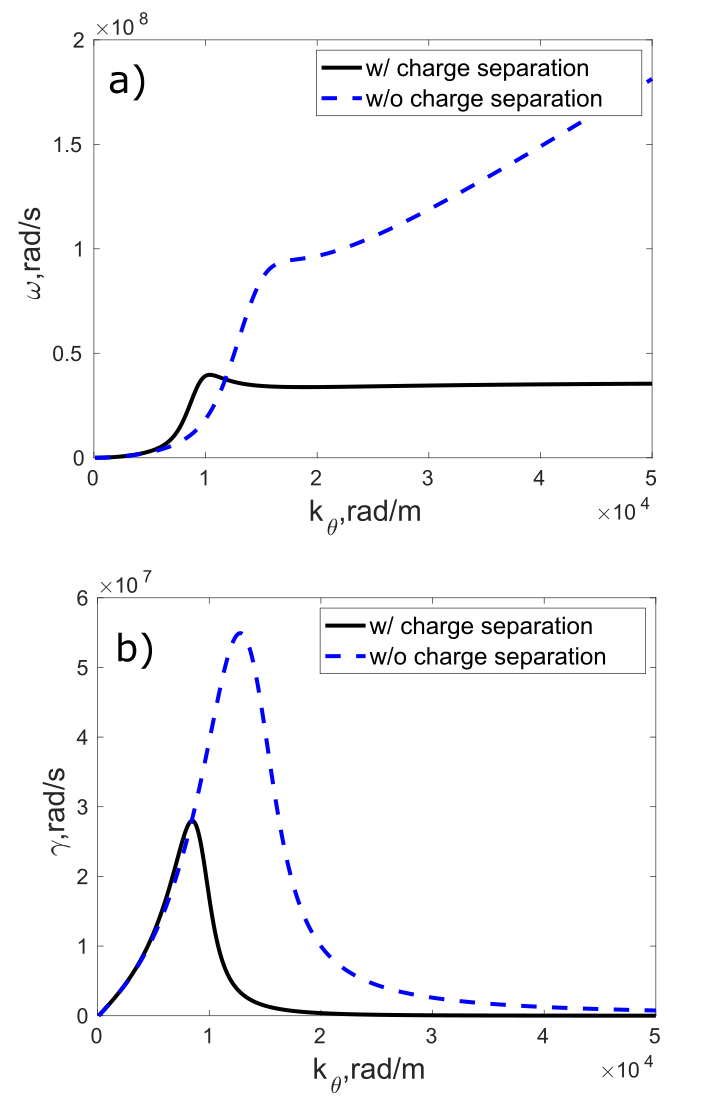}
\caption{Theoretical prediction  of a) real frequency and b) growth rate when $B=40{\rm mT}$ with/without the charge separation effect using Eq. 1 and Eq. 2. The calculation was performed with $n_0=3 \times 10^{16} m^{-3}$, $T_e=5 {\rm eV}$, $\nu_{en}=2.5 \times 10^8 {\rm s^{-1}}$ and the other parameters input from Table 1, e.g., $E_a=2484{\rm V/m},L_n=0.51{\rm mm}$.}
\end{figure} 

In the PIC/MCC simulations shown later in the present work, the plasma density is about $10^{16} {\rm m^{-3}}$ so that the Debye length $\lambda_{De}$ can be comparable to the wavelength $1/k_{\theta}$, suggesting that the non-neutrality (charge separation) may play a role. With the  charge separation effect taken into consideration (see appendix for the derivation),  Eq. 1 is modified to:

\begin{equation}
(k_{\theta}\lambda_{De})^2=\frac{k^2_{\theta}c^2_s}{\omega^2} - \frac{\omega_{d}+k^2_{\theta}\rho^2_{e}(\omega-\omega_{E}+i\nu_{en})}{\omega-\omega_{E}+k^2_{\theta}\rho^2_{e}(\omega-\omega_{E}+i\nu_{en})},
\end{equation}

The term at the left hand introduces the charge separation effect. One example showing the real frequency and the growth rate obtained from Eq. 1 and Eq. 2 is given in Fig. 1. The input parameters in the equations for plots are from Table 1 in Sec. 5 when $B=40 {\rm mT}$. In this particular case, $\rho_e=0.13 {\rm mm}$ and $\lambda_{De}=0.096 {\rm mm}$, meaning $ k_{\theta}\rho_{e} \approx k_{\theta}\lambda_{De} \approx 1$ when $k_{\theta} \approx 1.0 \times 10^4 {\rm rad/m}$. As shown in Fig. 1a, different modes of the gradient drift instability with the assumption of quasineutrality are presented: Simon-Hoh modes when $k_{\theta} \lesssim 5000 {\rm rad/m}$ ($k_{\theta}\rho_e \lesssim 0.5$); the ion sound modes when $k_{\theta} \gtrsim 20000 {\rm rad/m}$ ($k_{\theta}\rho_e \gtrsim 2$); lower hybrid modes in between. With the charge separation effect considered, the real frequency and the growth rate are fairly altered in the ion sound modes and the lower hybrid modes. The ion sound instability turns out to be a pure oscillation with the ion plasma frequency $\omega_{pi}=3.6\times10^7 {\rm rad/s}$ when $k_{\theta}>10^4 {\rm rad/m}$ ($k_{\theta}\lambda_{De} > 1$). Accordingly, the lower hybrid instability shifts to the longer wavelength regime, and the peak values of the real frequency and the growth rate become smaller. As expected, the real frequency and the growth rate in the long wavelength Simon-Hoh instability are less affected. 

In this work, we consider a crossed field discharge where two concentric cylindrical electrodes is applied by a constant voltage to drive the magnetized plasma with uniform axial magnetic field. By applying the above theory with the cylindrical geometry, the magnetic field is in the axial direction ($\mathbf{z}$ direction), the density gradient and the self-consistent electric field are directed radially ($\mathbf{x}$ direction) and the instability develops in the azimuthal direction ($\mathbf{y}$ direction). Instead of the linear perturbation of quantities $\sim exp[-i(\omega t - k_{\theta} y)]$ in the Cartesian coordinate, the harmonics $\sim exp[-i(\omega t-m\theta)]$ are sought in the cylindrical coordinate. Here, $m$ is the azimuthal mode number and correlates the wavenumber with $m=k_{\theta}r_0$ where $r_0$ is a radial position where the instability is excited.

\section{PIC/MCC model description}

The present work was motivated by the magnetically enhanced hollow cathode arc discharge (ME-HCAD) experiments \cite{fietzke2009magnetically,fietzke2010plasma,zimmermann2011spatially}. ME-HCAD generates extremely high electron density plasma inside the hollow cathode, together with the formation of a large area plasma plume for surface modifications, e.g., the film deposition and the sputter etching. Our model was performed to mimic the plasma core region of ME-HCAD with the typical experimental parameters \cite{fietzke2009magnetically}. As shown in Fig. 2, the model describes a hollow cathode with radius $r_2=6 {\rm mm}$ and with the uniform magnetic field directed outward in the axial direction. The center anode is applied here to simulate the one that is concentrically arranged with the hollow cathode in the experiments. The anode radius is $r_1=1 {\rm mm}$ and the electrode separation is $d = 5 {\rm mm}$, presenting the simulation box size. The magnetic field intensity is in the range of $10-60 {\rm mT}$, in which electrons are magnetized ($\rho_e \ll d$), while ions are not (ion Larmor radius $\rho_{i} \gg d$).  A voltage $U=200 {\rm V}$ typical to the experiments is applied between the electrodes to drive the discharge and the cathode is grounded. The self-consistently generated electric field in the radial direction results in the $\mathbf{E_r} \times \mathbf{B}$ configuration and the azimuthally closed electron drift. The working gas is argon, the pressure is $P=10 {\rm Pa}$, corresponding to the gas flow rate $10 {\rm sccm}$ in the experiments, and the background neutral density is assumed constant at the room temperature. The ME-HCAD plasma is generated and sustained by cathode-emitted thermionic electrons. The emission current density is up to $j_{emit}=5\times10^4 {\rm A/m^2}$, resulting in the plasma density around $10^{20} {\rm m^{-3}}$ and the Debye length $\lambda_{De}\sim 0.5 {\rm \mu m}$ that PIC cannot resolve. This challenge is usually dealt with by scaling the system in some ways, e.g., by reducing device size, decreasing the plasma density or increasing the relative permittivity (to increase the cell size). The present work adopted the method of reduced plasma density, meaning that the emission current density $j_{emit}$ is reduced. With the emissive electron injection, the steady state can be achieved efficiently as the previous study verified \cite{Boeuf2020new}. The quasi steady state is achieved after a runtime of $t \approx 1.5 {\rm \mu s}$ in our simulations.

\begin{figure}
\center
\includegraphics[clip,width=0.4\linewidth]{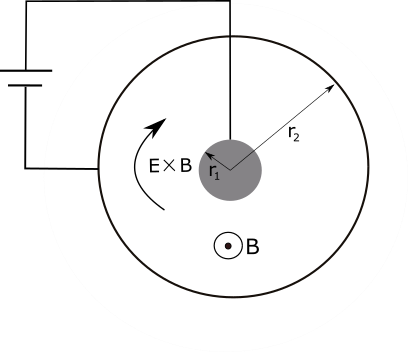}
\caption{2D radial-azimuthal PIC/MCC model of the crossed field plasma with cathode radius $r_2=6 {\rm mm}$ and anode radius $r_1=1{\rm mm}$. The gas pressure is $P=10 {\rm Pa}$ and the uniform axial magnetic field directs outward in the axial direction. }
\end{figure}

A two-dimensional radial-azimuthal explicit and electrostatic particle-in-cell code \cite{birdsall1991} was utilized to investigate the instability development and spoke dynamics. The code was extensively benchmarked and validated \cite{Jo2018,ganesh2002formation,ganesh2002dynamics} and the convergence was also verified with different time steps, cell sizes and particles per cell in our study. An equidistant computational grids $64 \times 512$ in the radial and azimuthal dimensions was used. The cell size is therefore $\Delta r=5 \times 10^{-5} {\rm m}$ in the radial direction, and the maximum cell size in the azimuthal direction is $\Delta l_{\theta} = 5 \times 10^{-5} {\rm m}$. Initially, the total number of computational particles in the simulation box is about $5\times 10^6$ and the particles are distributed that the plasma density is roughly uniform $n_{e0}=n_{i0}=1 \times 10^{16} {\rm m^{-3}}$. The number of particles per cell is $N_p=140$ initially and can be about $N_p=300$ in the steady state. The electron temperature is set to be $T_e=5 {\rm eV}$ and ions have room temperature $T_i=0.026 {\rm eV}$.

The value of the electron current density produced by the thermionic emission $j_{emit}\approx 0.1-0.6 {\rm A/m^2}$ is set and adjusted at different magnetic fields to obtain the peak plasma density in the steady state to be roughly $5\times 10^{16}{\rm m^{-3}}$. This ensures that the Debye length can be resolved with the given cell sizes $\Delta r$ and $\Delta l_{\theta}$. The time step is set to be $\Delta t = 10 {\rm ps}$ to resolve the smallest time scale of electron plasma frequencies. In our simulations, ions are assumed to be collisionless due to the long ion-neutral mean free path $\lambda_i \sim d$ and electron-neutral collisions are tracked by the Monte Carlo method. The elastic and ionization collisions are considered, and the excitation is neglected. The electron-neutral collision cross sections are those of Phelps \cite{phelps1999cold}.  Based on the simulation setup, the ion sound velocity is $c_s=3.5 {\rm km/s}$, electron-neutral collision frequency is $\nu_{en} \approx 2.5 \times 10^8 s^{-1}$, the Debye length is approximated to be $\lambda_{De} = 0.1 {\rm mm}$ and  $\rho_{e} = 0.13 {\rm mm}$ when $B=40 {\rm mT}$. Under these conditions, the effects of gyro-viscosity and non-neutrality are important, particularly in the short wavelength regime ($k_{\theta}\rho_e > 1$, $k_{\theta}\lambda_{De} \gtrsim 1$).

The code is accelerated by OpenMP parallelization and the simulation was carried on a 32-processor Intel Xeon workstation, with a run duration of about 10 days. About $1.5\times 10^6$ time steps were performed, corresponding to a simulated time of $15 {\rm \mu s}$, during which the spoke rotates about two periods.

\section{Observation of a rotating spoke}

In this section, the rotating spoke/macroscopic structure phenomena are shown in the saturated nonlinear stage. The cases shown have the magnetic field induction of $B=60 {\rm mT}$ and $B=20 {\rm mT}$. Other simulation parameters, such as the discharge pressure, the anode applied voltage, the electrode separation stated in Sec. 3, are kept fixed through this work. 
 
 \begin{figure}
\center
\includegraphics[clip,width=0.9\linewidth]{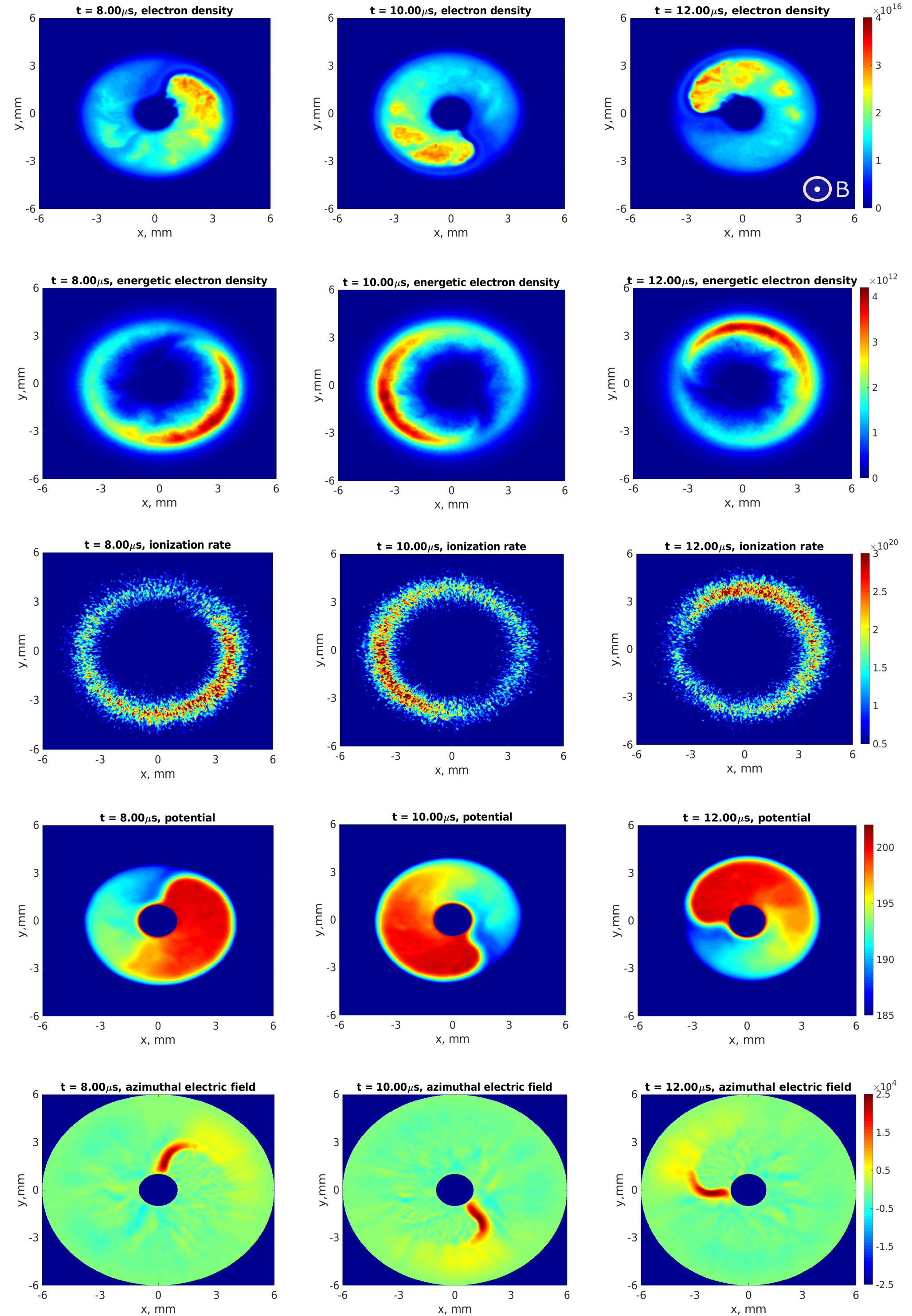}
\caption{Evolution of the total electron density $n_e$, the energetic electron density $n_f$ (with energy $\varepsilon>15.78 {\rm eV})$, the ionization rate $\sigma_{ioniz}$, the potential $\phi$ and the azimuthal electric field $E_{\theta}$ in the saturated nonlinear stage showing the rotating spoke when $B=60{\rm mT}$. The m=1 spoke rotates as a rigid body in the $\mathbf{E_r}\times \mathbf{B}$ (clockwise) direction with the frequency of $0.16 {\rm MHz}$. }
\end{figure} 
 
\begin{figure}
\center
\includegraphics[clip,width=0.9\linewidth]{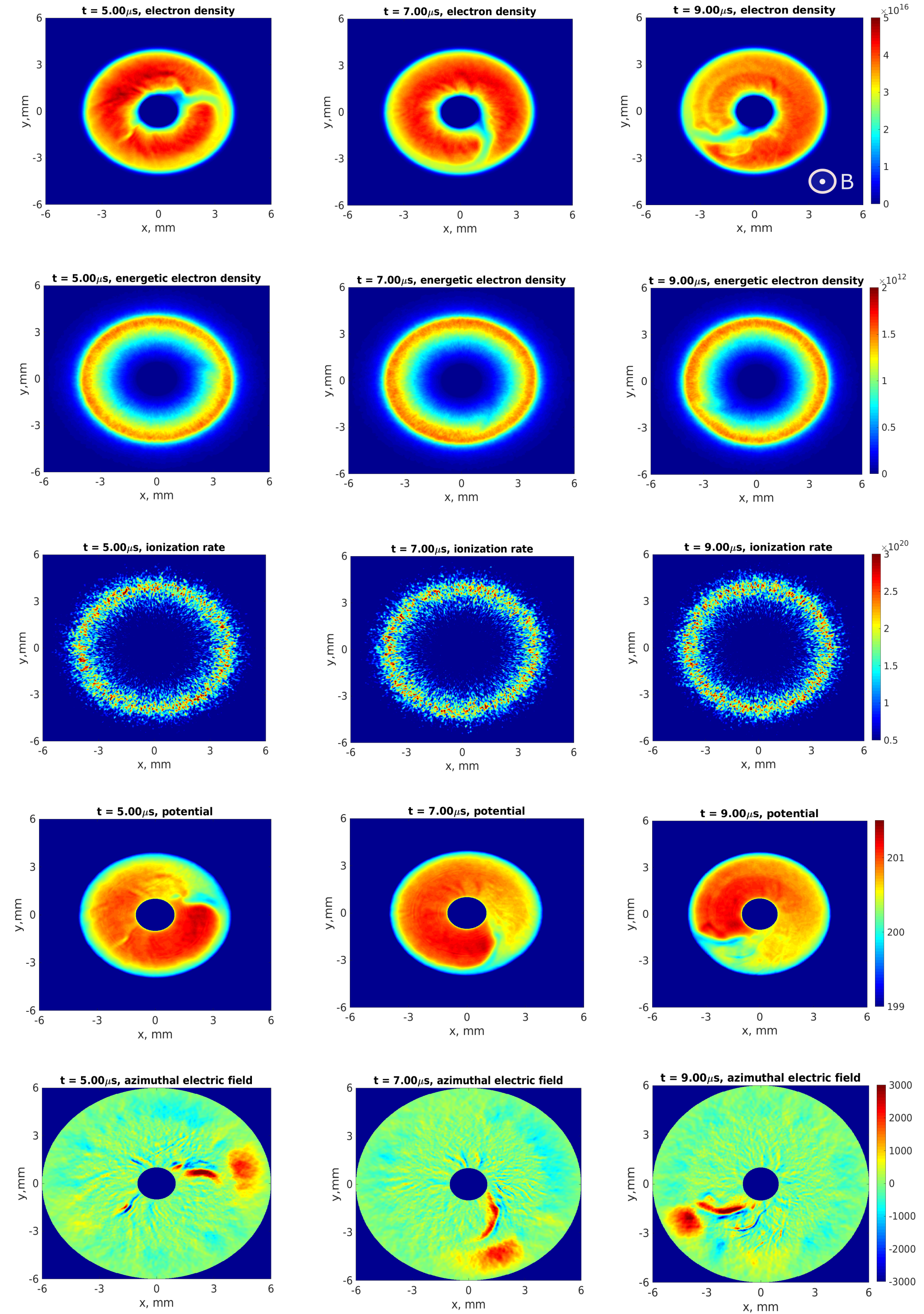}
\caption{Evolution of the electron density $n_e$, the energetic electron density $n_f$( with energy $\varepsilon>15.78 {\rm eV})$, the ionization rate $\sigma_{ioniz}$, the potential $\phi$ and the azimuthal electric field $E_{\theta}$ in the saturated nonlinear stage showing the rotating macroscopic structure when $B=20{\rm mT}$. The m=1 macroscopic structure rotates as a rigid body in the $\mathbf{E_r}\times \mathbf{B}$ (clockwise) direction with the frequency of $0.1 {\rm MHz}$. }
\end{figure} 
 
Fig. 3 shows the time evolution of the total electron density $n_e$, the energetic electron density $n_{f}$ (with energy $\varepsilon>15.76 {\rm eV}$), the ionization rate $\sigma_{ioniz}$, the potential $\phi$ and the azimuthal electric field $E_{\theta}$ in the steady state when $B = 60 {\rm mT}$. The spoke, referred to as the region of enhanced light emission/excitation, can be identified through the $\sigma_{ioniz}$ plot (The excitation rate exhibits the same feature when we included the excitation collision). The spoke rotates as a rigid body in the $ \mathbf{E_r} \times \mathbf{B}$ (clockwise) direction with the frequency $f=0.16{\rm MHz}$ and the velocity $v_s=4.0 {\rm km/s}$ given $ r = 4 {\rm mm}$. The characteristics are consistent with various experimental observations \cite{Anders2012,Anders2017,McDonald2011,Hecimovic2018,Ito2015,Marcovati2020,parker2010}.  From Fig. 3, it can be seen the enhanced ionization is associated with the energetic electron concentration, indicating electron heating is enhanced there. From the plot of $E_{\theta}$, the spoke front of the double layer nature (see the potential plot), with magnitude of $E_{\theta} \approx 25 {\rm kV/m}$ is seen. It is therefore suggested that the electron heating are associated with the $\mathbf{E_{\theta}} \times \mathbf{B}$ electron flow channeled by the the spoke front passage (see discussions in Sec. 6.2). With regard to the electron density $n_e$, the plasma is also self-organized as the well-shaped structure rotating in the same manner as the spoke. Therefore, in the case of $B=60 {\rm mT}$, the well-established rotating spoke can be observed.

Likewise, Fig. 4 shows the same plots when $B = 20 {\rm mT}$. In this case, the profiles of $\sigma_{ioniz}$ and $n_f$ do not exhibit the localized enhanced ionization events and the enhanced energetic electron concentration. The spoke scenario is therefore hardly observed by a naked eye. While the well-shaped m=1 large scale structure can still be seen in terms of $n_e$. The structure rotates in the $ \mathbf{E_r} \times \mathbf{B}$ direction with the frequency $f=0.1 {\rm MHz}$ and the velocity $v_s=2.8 {\rm km/s}$ given $r=4 {\rm mm}$. It is noteworthy the double layer in front of the large structure is also present as shown in the $\phi$ plot. The magnitude of $E_{\theta}$ in the double layer is about $2.5 {\rm kV/m}$, suggesting that the $\mathbf{E_{\theta}} \times \mathbf{B}$ electron flow associated heating does not suffice to provide the enhanced ionization to support the spoke formation. Here, we refer to the large scale structure as the rotating macroscopic structure when B is so low that the enhanced ionization is not visible.

We also performed simulations for different magnetic fields. When $B=10 {\rm mT}$, the system is stable and no macroscopic structure forms. When B is increased from $20 {\rm mT}$ to $60 {\rm mT}$, the magnitude of spoke front $E_{\theta}$ becomes larger, resulting in the stronger electron heating and hereby more pronounced spoke. 

To summarize, the spoke/macroscopic structure forms given the instability is excited. The electron heating does not suffice to support the formation of spoke when $B = 20 {\rm mT}$, while the heating is enhanced with $B$ increasing until the spoke is well established when $B = 60 {\rm mT}$.

 \begin{figure}
\center
\includegraphics[clip,width=0.5\linewidth]{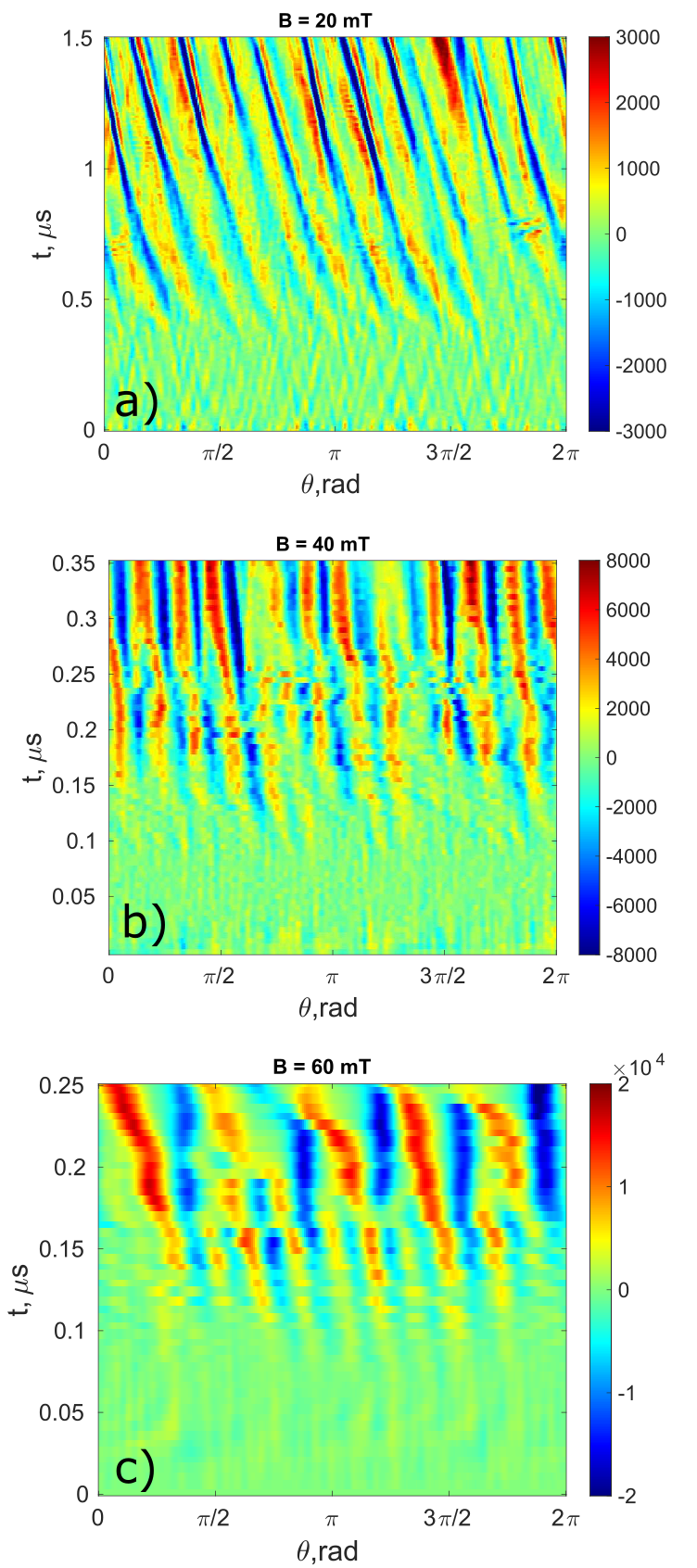}
\caption{The spatial and temporal evolution of $E_{\theta}$ showing the instability development when a) $B=20 {\rm mT}$, b) $B=40 {\rm mT}$ and c) $B=60 {\rm mT}$.}
\end{figure}

\begin{figure}
\center
\includegraphics[clip,width=0.4\linewidth]{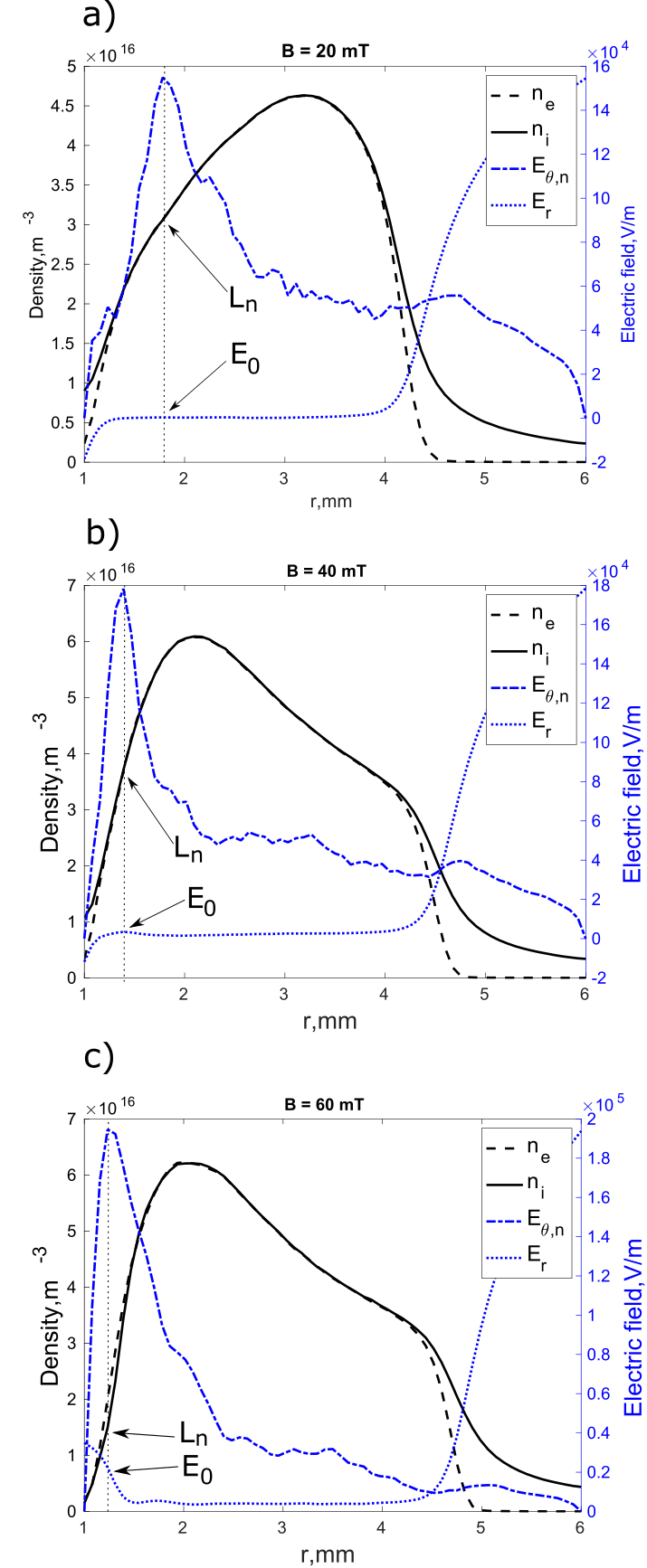}
\caption{Radial profiles of the azimuthally averaged electron density $n_e$, the ion density $n_i$, the radial electric field $E_r$ and the azimuthal electric field $E_{\theta,n}$ in the linear stage a) at the snapshot $t=0.56\mu s$ when $B=20 {\rm mT}$, b) at $t=0.16 \mu s$ when $B=40{\rm mT}$ and c) at $t=0.16 \mu s$ when $B=60 {\rm mT}$. The vertical lines denote the radial positions where $E_{\theta,n}$ are peaked, i.e., the instability is excited. $L_n$ is the local density gradient and $E_0$ is the radial electric field at the position $r=r_0$.}
\end{figure}

\begin{figure}
\center
\includegraphics[clip,width=0.9\linewidth]{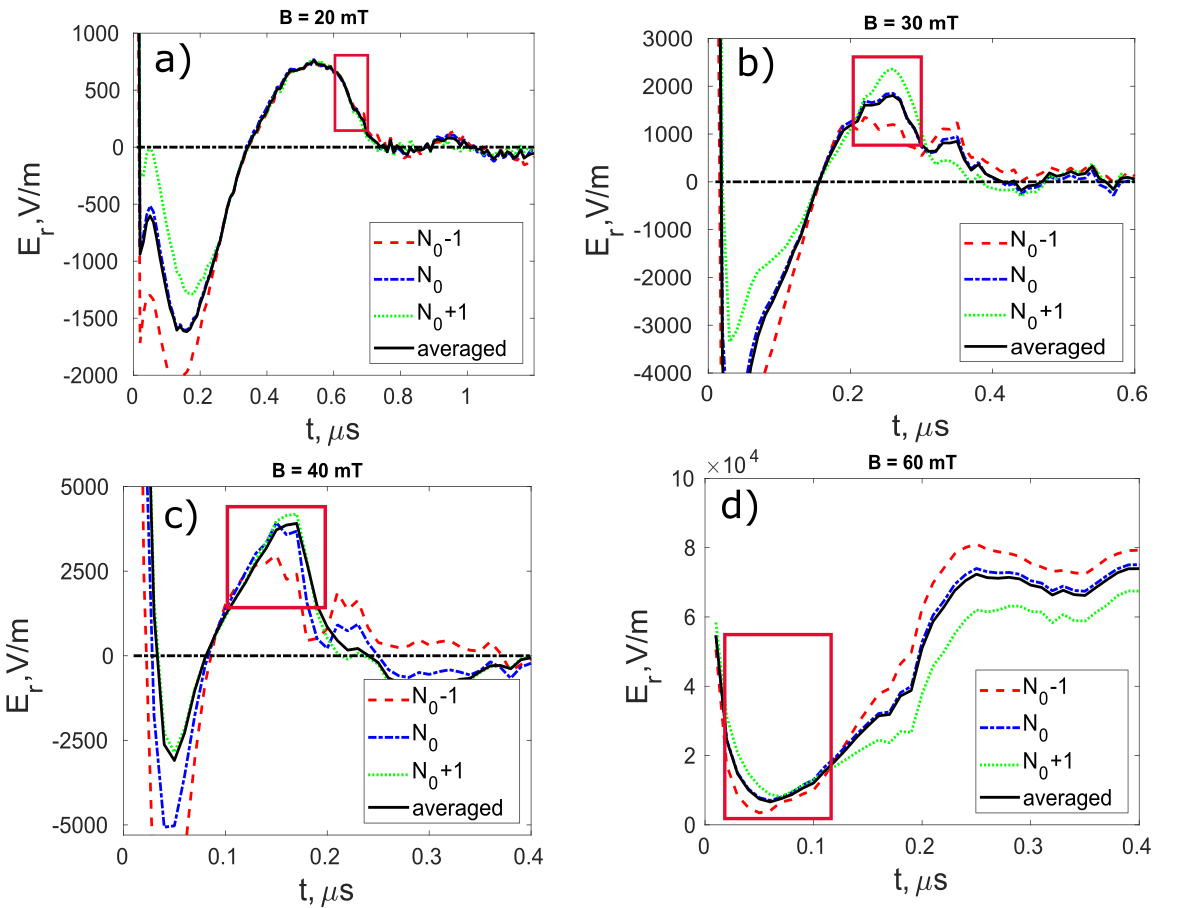}
\caption{Time history of azimuthally averaged radial electric field at three adjacent cell grids $N_0-1$, $N_0$ and $N_0+1$ when a) $B=20 {\rm mT}$, b) $B=30{\rm mT}$, c) $B=40 {\rm mT}$ and d) $B=60 {\rm mT}$. $N_0$ is the cell number corresponding to the radial location $r=r_0$. At each moment, $E_r$ is further averaged over the three consecutive grid points to generate the black solid line at different $B$. The red box represents the time period, in which the fluctuation grows linearly in a smooth manner (see Fig. 9). The time period is $[0.6\mu s, 0.7 \mu s]$ for $B=20 {\rm mT}$,  $[0.2\mu s, 0.3 \mu s]$ for $B=30 {\rm mT}$,  $[0.1\mu s, 0.2 \mu s]$ for $B=40 {\rm mT}$,  $[0.02\mu s, 0.12 \mu s]$ for $B=60 {\rm mT}$. $E_r$ (black solid line) inside the box is further averaged in time to obtain the totally-averaged (radially+azimuthally+in time) radial electric field $E_a$, together with the maximum value $E_{max}$ and the minimum value $E_{min}$ within the time period (see Table 1 for their values).}
\end{figure}

 \begin{figure}
\center
\includegraphics[clip,width=0.5\linewidth]{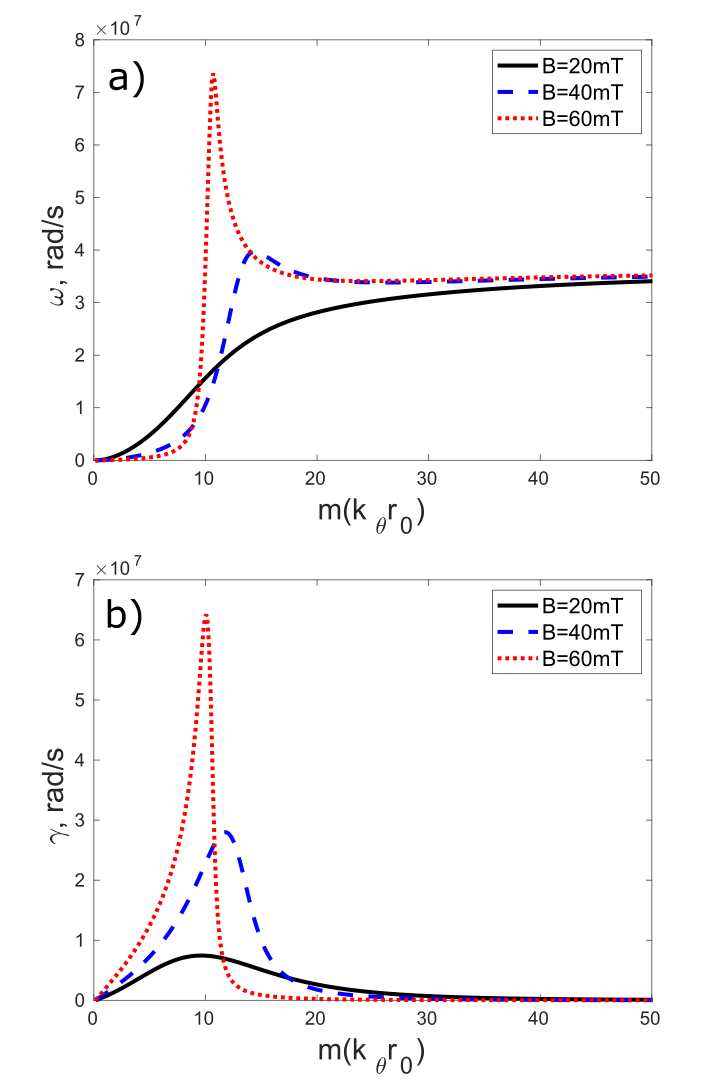}
\caption{Theoretical prediction of a) the real frequency and b) the grow rate for different magnetic fields using Eq. 2 with $n_0=3 \times 10^{16} {\rm m^{-3}}$, $\nu_{en}=2.5 \times 10^8 {\rm s^{-1}}$, $T_e=5 {\rm eV}$ and the values of $E_a$ and $L_n$ shown in Table 1.}
\end{figure}

\section{Identification of instabilities in the linear stage}

Following the observation of the spoke/macroscopic structure in the nonlinear stage, we backtraced the micro instabilities in the linear stage. In this section, the instability features, i.e., the growth rate and the $\omega-k_{\theta}$ spectrum, were obtained to identify the instability by comparing them to the theory described by Eq. 2. The comparison was performed across the magnetic field range $B=10-60 {\rm mT}$, and the electron Larmor radius varies from $\rho_e=0.27{\rm mm}$ when $B=20 {\rm mT}$ to $\rho_e=0.089 {\rm mm}$ when $B=60 {\rm mT}$.


The spatial and temporal evolution of $E_{\theta}$ presenting the instability development is shown in Fig. 5 when $B=20 {\rm mT}$, $B=40 {\rm mT}$ and $B=60 {\rm mT}$. The data is taken at the radial position $r = r_0$ where the instability is excited. The values of $r_0$ are listed in Table 1. From Fig. 5, the instability can be seen by a naked eye and the mode number $m$ can be roughly estimated. The precise value of $m$ is derived in the Fourier space and given in Table 1. 

According to Eq. 2, the density gradient length $L_n$ and the radial electric field $E_r$ are required to obtain the theoretical growth rate and real frequency. Fig. 6 shows the radial profiles of azimuthally averaged electron density $n_e$, ion density $n_i$, radial electric field $E_r$ and normalized azimuthal electric field $E_{\theta,n}$ at snapshots in the linear stage of the instability for $B=20 {\rm mT}$, $B=40 {\rm mT}$ and $B=60 {\rm mT}$. To make the azimuthal electric field visible in Fig. 6, it is scaled as

\begin{equation}
  E_{\theta,n}(r,t)=\sqrt{\int_0^{2\pi} E_{\theta}(r,\theta,t)^2 d\theta/2\pi}\frac{\max(\int_0^{2\pi} E_{r}(r,\theta,t) d\theta/2\pi)}{\max(\sqrt{\int_0^{2\pi} E_{\theta}(r,\theta,t)^2 d\theta/2\pi})},   
\end{equation}

From the profile of $E_{\theta,n}$, the radial position $r_0$ where the instability is excited can be accurately located according to the $E_{\theta,n}$ peak value. As shown in Fig. 6 for all cases, the instability is initiated close to the anode where the density gradient is large, indicating its important role for the instability development. The values of $L_n$ and $E_0$ at the position $r_0$ can be precisely obtained. We found $L_n$ rarely changes radially and with time in the linear stage of the instability, while $E_0$ is sensitive to the time evolution. To estimate the spatial non-locality and follow the time evolution, the azimuthally averaged $E_r$ is plotted as a function of time and at three adjacent cell grids ($N_0-1, N_0, N_0+1$) for different magnetic field cases in Fig. 7. $N_0$ is the grid number corresponding to the position $r=r_0$. The red box represents the time period, during which the linear fluctuation growth is fitted to obtain the growth rate (see Fig. 9). As shown in the red box, overall, the fluctuation of $E_r$ in time is much stronger than that in the radial positions, particularly for $B=60 {\rm mT}$ case. The black solid curve gives $E_r$ radially averaged over the three adjacent grid points at each moment. Inside the red box, $E_r$ is further averaged in time to obtain the totally-averaged radial electric field (radially, azimuthally and in time), i.e., $E_a$, and the maximum value $E_{max}$ and minimum value $E_{min}$ during the time period. $E_a$ is applied as input for the theoretical prediction of the real frequency and the growth rate, and $E_{max}, E_{will}$ are used to estimate the uncertainty. The values of $E_a$, $E_{max}$, $E_{min}$ and $L_n$ are listed in Table 1. From Fig. 7, it is very interesting to point out that the instability occurs only when $E_r$ is positive, meaning $\mathbf{E_r} \cdot \triangledown n_0>0$, that is characteristic of the Simon-Hoh instability.   

With the input of $E_a$ and $L_n$ in Eq. 2, the theoretical real frequency and growth rate are derived and plotted in Fig. 8 for different magnetic fields. Due to the charge separation effect, as expected for all cases, the short wavelength ion sound instability becomes purely oscillatory with the frequency $\omega_{pi}\approx 3.6 \times 10^7 {\rm rad/s}$ as shown in Fig. 8a. From Fig. 8b, the short wavelength instability perturbations when $m>30$ are stabilized due to the electron inertia effect \cite{lakhin2018effects}. Besides, the electron inertia results in the occurrence of perturbations with the frequency on the order of lower hybrid frequency $\omega_{lh}=\sqrt{\omega_{ce}\omega_{ci}}$, where $\omega_{ci}$ is the ion cyclotron frequency. For example, in our cases, $\omega_{lh}=3.9 \times 10^7 {\rm rad/s}$ for $B=60 {\rm mT}$ and $\omega_{lh}=2.6 \times 10^7 {\rm rad/s}$ for $B=40 {\rm mT}$ approximately correspond to the peak values of the non-monotonous real frequency as shown in Fig. 8a. For $B=20 {\rm mT}$, $\omega_{lh}=1.3 \times 10^7 {\rm rad/s}$ is smaller than $\omega_{pi}$, yielding the monotonous real frequency as a function of the mode number. From Fig. 8b, the theoretical growth rate $\gamma_{t}$ at a certain mode can be precisely obtained. The most unstable mode (with the largest growth rate) can also be determined and it is clearly seen the most unstable mode are in the lower hybrid instability from Fig. 8a and 8b. As shown later, the unstable modes developing in the simulations match well the lower hybrid instability.

The simulated growth rate can be obtained by fitting the curve of ion density fluctuation as a function of time in Fig. 9. The ion density fluctuation at the radial position $r_{0}$ is estimated as follows: 

\begin{equation}
\frac{\Delta n_{i}}{n_{0}}(r=r_0,t)=\frac{[\int^{2\pi}_0 n_i(\theta,t)^2 d \theta/2\pi-(\int^{2\pi}_0 n_i(\theta,t) d\theta/2\pi)^2]^{1/2}}{n_0},
\end{equation}

\noindent where $n_0=\int^{2\pi}_0 n_i d\theta/2\pi$ is the azimuthally averaged ion density at $r=r_0$. In Fig. 9, the logarithm coordinate for the y axis is applied to identify the linear growth and calculate the growth rate. As stated, the linear stage and nonlinear saturation can be clearly identified in Fig. 9, providing the possibility to separate the linear stage for the precise analysis. 

\begin{figure}
\center
\includegraphics[clip,width=0.7\linewidth]{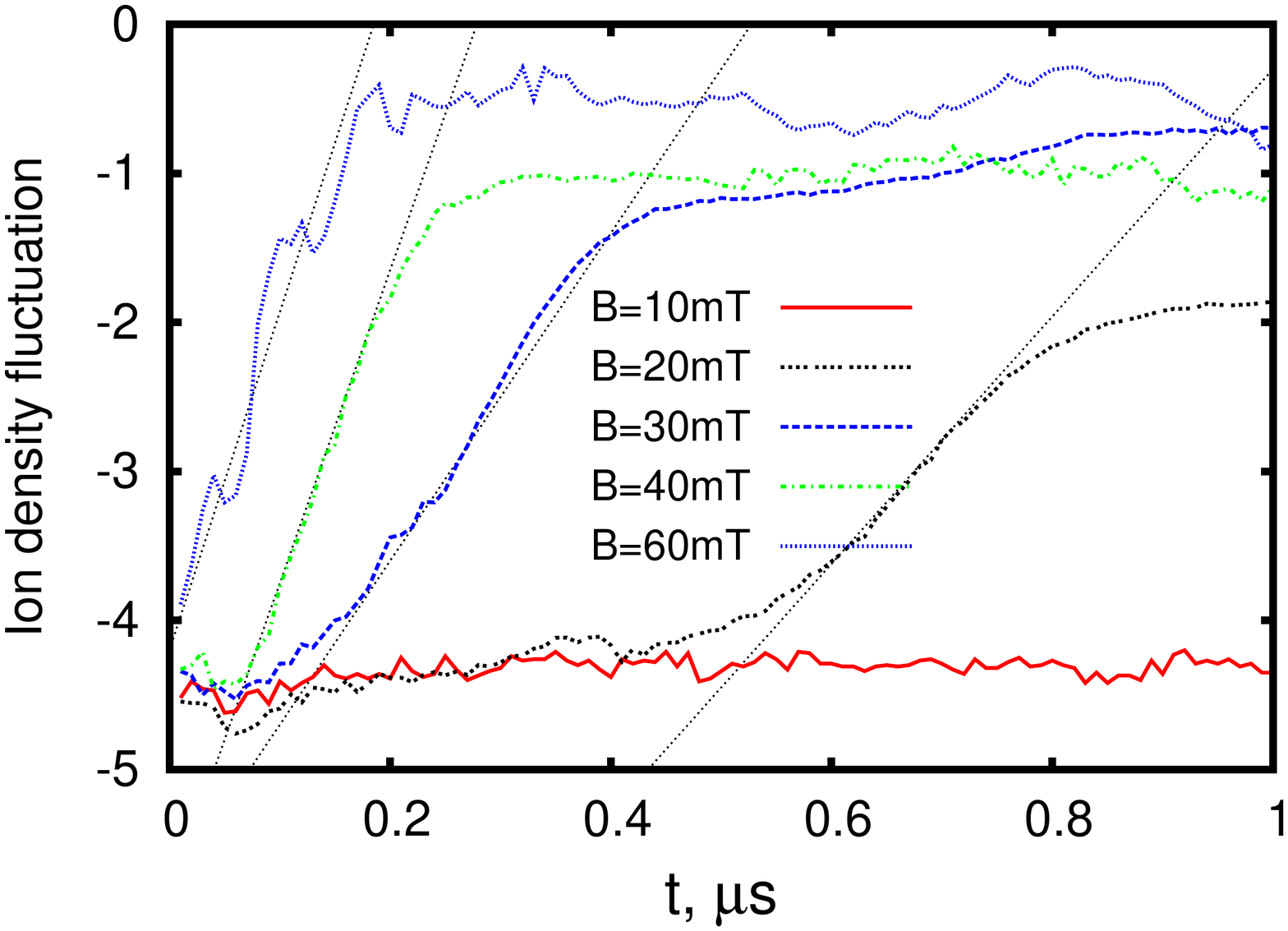}
\caption{Ion density fluctuation as a function of time for different magnetic fields. The fluctuation is shown in logarithm coordinate.}
\end{figure} 

\begin{table}[]
    \centering
\hspace*{-1.4cm}\begin{tabular}{ |p{1.2cm}||p{1.3cm}|p{1.5cm}|p{1.9cm}|p{1.9cm}|p{1.4cm}|p{0.5cm}|p{1.7cm}|p{1.5cm}|p{1.55cm}|}
 \hline
$B (mT)$ &  $r_{0}(mm)$ & $E_a (V/m)$ & $E_{max} (V/m)$ & $E_{min} (V/m)$ & $L_n (mm)$ &$m$ & $k_{\theta} (rad/m)$ & $\gamma_{t} (rad/s)$ & $\gamma_{s}(rad/s)$ \\
 \hline
20   &1.78  &422   &661  &111  & 1.77  & 10  &5614   &7.4e6  &8.2e6\\
30   &1.47  &1436  &1806 &776  & 1.06  &12   &8125   &10.1e6  &10.7e6\\
40   &1.39  &2484  &3911 & 1205 & 0.51  &12    &8622   &27.5e6  &20.7e6\\
60   &1.23  &14436 &38921 &7415 & 0.20  &10  &8132   &63.0e6 &23.2e6\\
 \hline
\end{tabular}
\hspace*{-1cm}\caption{The table lists the parameters obtained from the self-consistent PIC/MCC simulations to calculate the instability dispersion relation using Eq. 2. $r_{0}$ is the radial position where the instability is initiated, $E_a$ is the totally-averaged radial electric field and $L_n$ is the density gradient length at the position $r_{0}$. For all cases at different magnetic fields, Debye length is $\lambda_{De}=0.096 {\rm mm}$ by using the local plasma density $n_0=3 \times 10^{16}{\rm m^{-3}}$ and the electron temperature $T_e=5eV$.  $\gamma_s$ is the instability growth rate obtained from the simulations and $\gamma_{t}$ is the theoretical growth rate using Eq. 2 with $E_a$ and $L_n$.}
\end{table}

\begin{figure}
\center
\includegraphics[clip,width=0.7\linewidth]{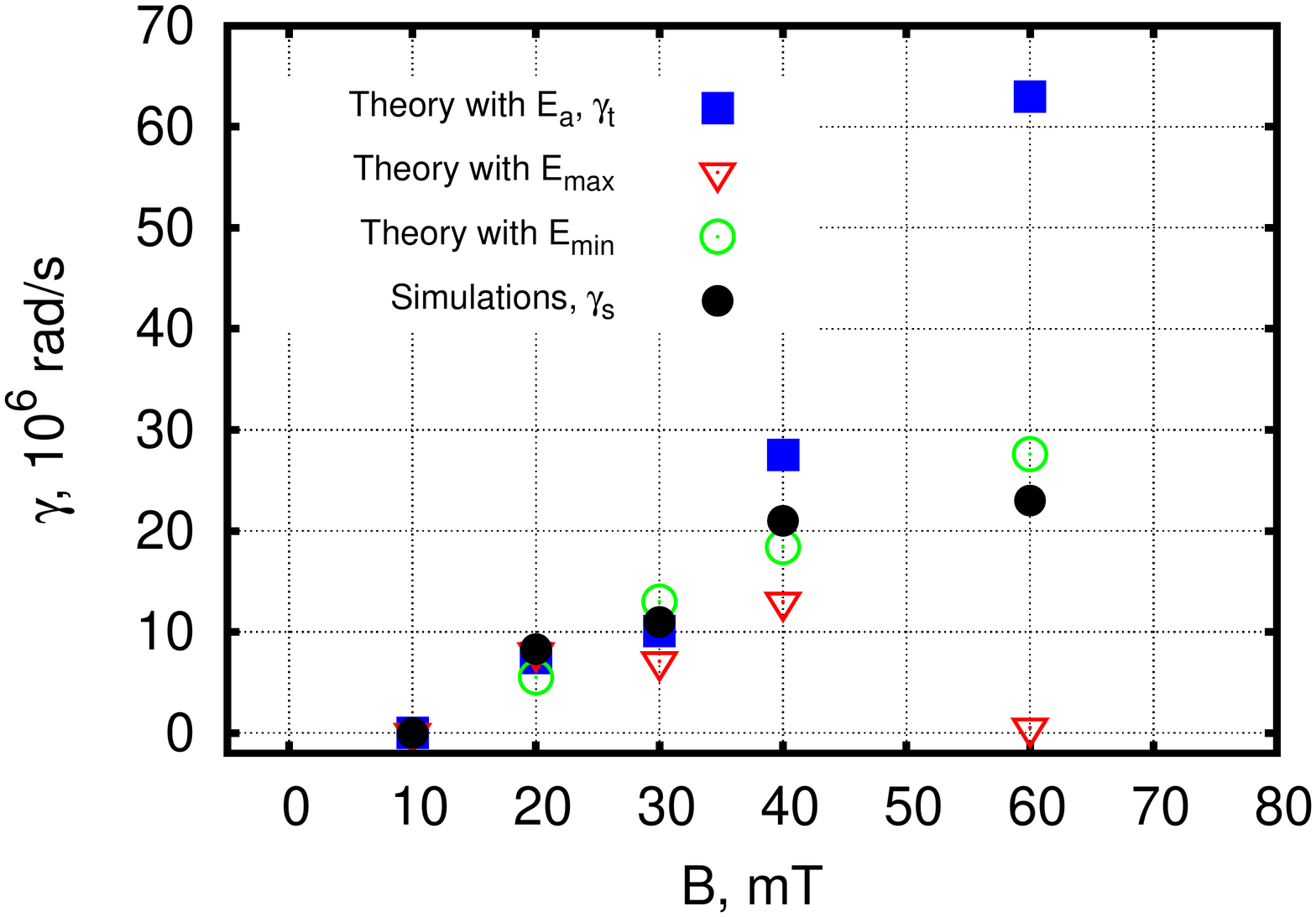}
\caption{Comparison of growth rates between simulations ($\gamma_s$) and theoretical calculations (using $E_a$, $E_{max}$ and $E_{min}$) as a function of magnetic field.}
\end{figure}

The simulated growth rates $\gamma_s$ are presented in Table 1 and compared to the theoretical values in Fig. 10. The theoretical predictions with $E_a$, $E_{max}$ and $E_{min}$ are all presented. It is seen that $\gamma_s$ (by using $E_a$) agrees very well with $\gamma_{t}$ in the magnetic range of $B=10-40 {\rm mT}$. Even using $E_{max}$ and $E_{min}$, the simulations still show very good agreement with the theory in the range of $B=20-40 {\rm mT}$. This may explain the smooth growth of the ion density fluctuation at these magnetic fields in Fig. 9. We note that the growth rate is zero when $B=10 {\rm mT}$, because the radial electric field near the anode is negative and the instability is not excited. There exists a large deviation between $\gamma_t$ and $\gamma_s$ when $B=60 {\rm mT}$. It can be attributed to the large oscillation of $E_r$ in the linear time period, which may also explain the oscillation in the ion density fluctuation growth when $B=60 {\rm mT}$ (see Fig. 9). The other possibilities are the formation of the ion flow accelerated in the large radial electric field and the collision frequency uncertainty \cite{smolyakov2016,xu2020self}.

 \begin{figure}
\center
\includegraphics[clip,width=0.5\linewidth]{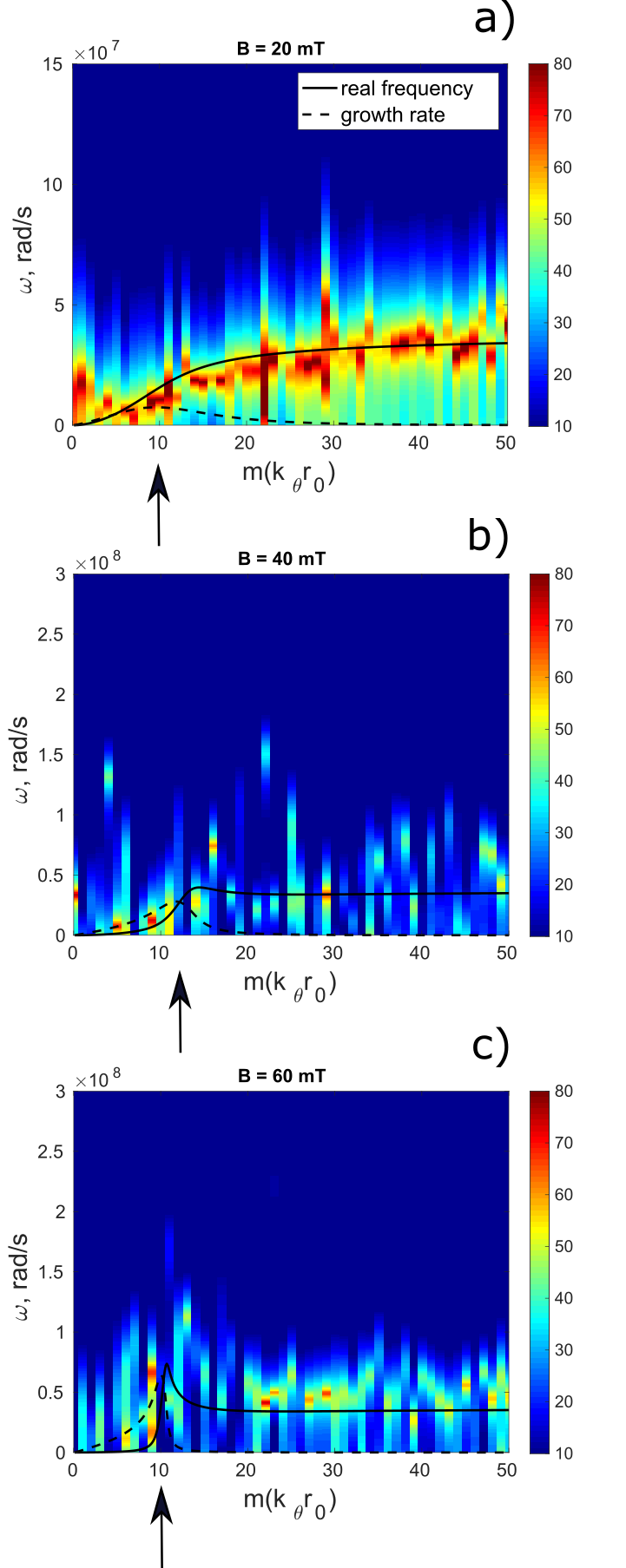}
\caption{FFT+MUSIC transform of the time-space data of $E_{\theta}$ in the linear stage shows the spectrum of the instability perturbation (see Fig. 4 for the $E_{\theta}$ data at different magnetic fields). The theoretical real frequency is plotted to compare the simulated spectrum and the growth rate is plotted to show the most unstable mode. The arrows designate the unstable modes developed in the simulations.}
\end{figure}

To further verify and confirm the instability, the $\omega-k_{\theta}$ spectrum by transforming the $\theta-t$ data of $E_{\theta}$ (see Fig. 4) is compared to the theoretical real frequency (see Fig. 8a). For standard fast Fourier transform (FFT) methods, the frequency resolution $\delta f$ is inversely proportional to the signal length. Thus, the duration of the instability's linear stage gives $\delta f \approx 1\times 10^6 {\rm Hz}$ when $B=20 {\rm mT}$, $\delta f \approx 4\times10^6 {\rm Hz}$ when $B=40 {\rm mT}$ and $\delta f \approx 5 \times 10^6 {\rm Hz}$ when $B=60 {\rm mT}$. The frequency predicted by the theory shown in Fig. 8a is peaked at $\omega/2\pi \approx 1\times 10^7 {\rm Hz}$ and is comparable to $\delta f$ when $B=40 {\rm mT}$ and $B=60 {\rm mT}$, indicating that the FFT method can not resolve the frequency space. In this work, a super-resolution signal processing approach MUSIC (multiple signal classification) \cite{hayes1996,kleiber2021}, which is able to overcome the frequency resolution limit and drastically reduce the noise, is applied. On the other hand, the $k_{\theta}$ space can be well resolved by FFT method transforming the $\theta$ dimension. Therefore, the combination of FFT and MUSIC was adopted to obtain the $\omega-k_{\theta}$ spectrum. The procedure can be interpreted as: 1) the FFT transforms the $\theta$ dimension to the $k_{\theta}$ space; 2) in each $k_{\theta}$-channel, MUSIC transforms time series to the frequency space. We note MUSIC as a parametric method, is highly sensitive in finding frequencies but does not provide information about mode structures because it does not give the true power magnitude of the mode (see Ref. 51 for MUSIC details).

 \begin{figure}
\center
\includegraphics[clip,width=0.5\linewidth]{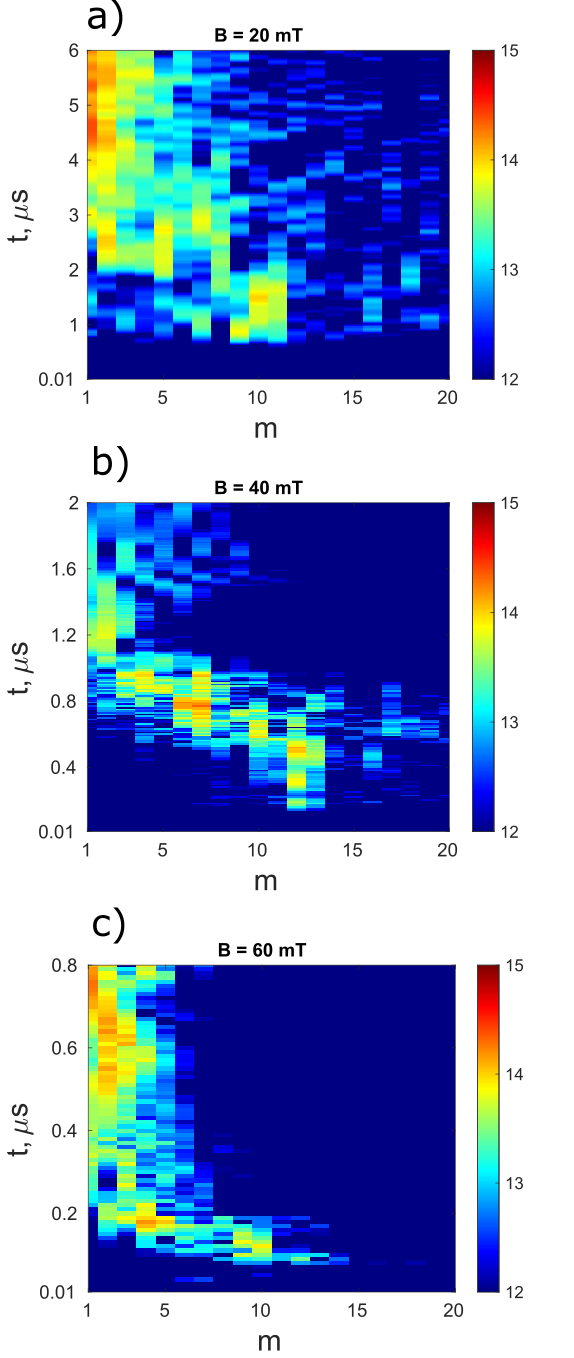}
\caption{The azimuthal electric field $E_{\theta}$ at $r=r_0$ in the azimuthal direction is transformed by the FFT method at each time step to obtain the $m-t$ plot for a) $B=20{\rm mT}$, b) $B=40{\rm mT}$ and c) $B= 60 {\rm mT}$. For all cases, the linear-nonlinear transition is accompanied by an inverse energy cascade.}
\end{figure} 

 \begin{figure}
\center
\includegraphics[clip,width=0.7\linewidth]{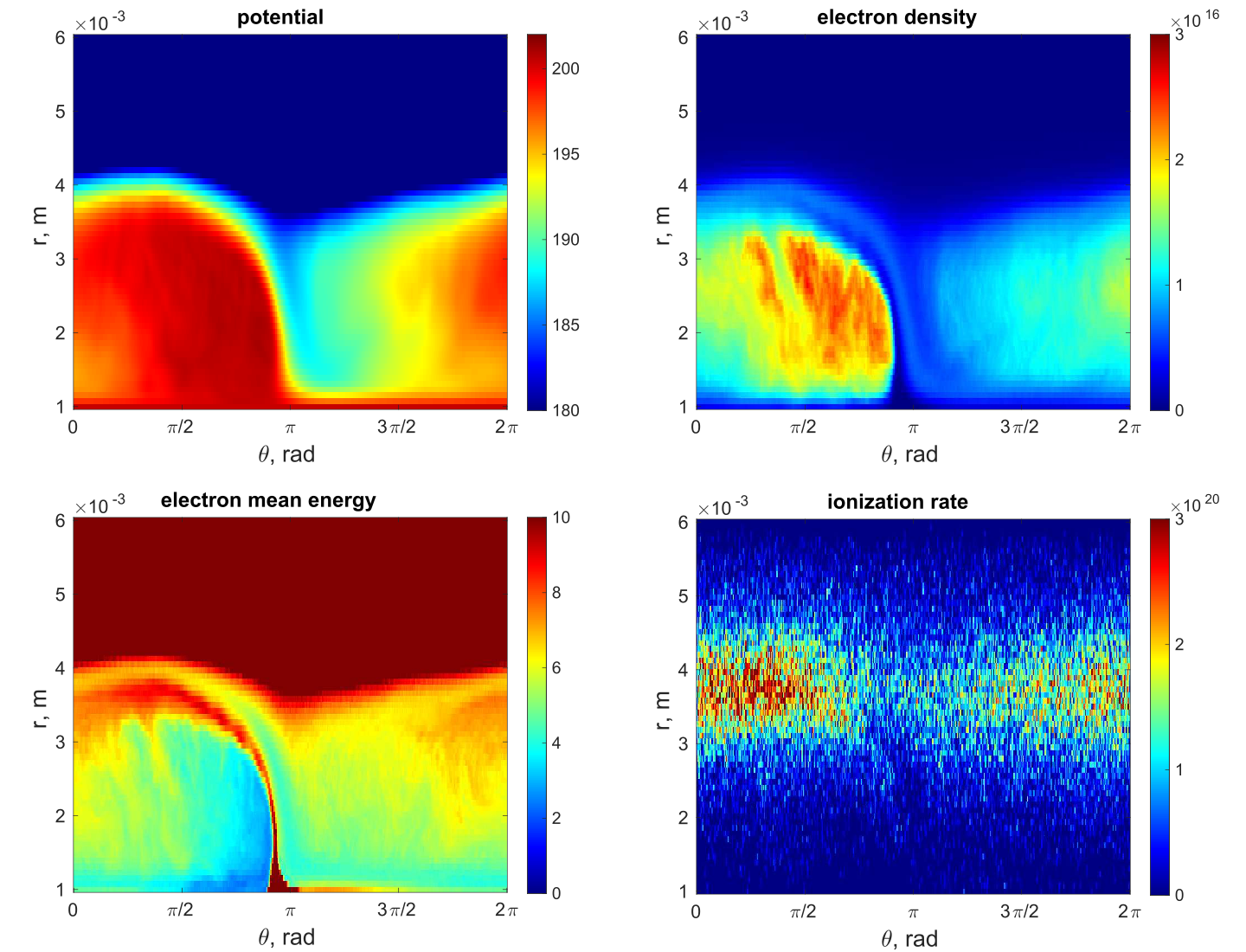}
\caption{a) $\theta-r$ profiles of a) potential $\phi$, b) total electron density $n_e$, c) electron mean energy $\varepsilon_m$ and c) ionization rate $\sigma_{ioniz}$ at the time snapshot $t=3.36 {\rm \mu s}$ for $B=60 {\rm mT}$.}
\end{figure}

The spectrum is displayed in Fig. 11, where the theoretical dispersion relation (both the real frequency and the growth rate) are also plotted for comparison. As shown in Fig. 11, the spectrum shows an excellent agreement with the theoretical real frequency for each case. It is seen that different modes of the gradient drift instability, i.e., Simon-Hoh, lower-hybrid and ion sound modes, are present in the spectrum for all the cases. Like the theory predicts, the ion sound instability is represented by the oscillation with $\omega_{pi}$ in the simulated spectrum. The arrows designate the linearly unstable mode developed in the simulations. We found that the developed modes agree well with the most unstable modes in the growth rate plot for the three cases. Therefore, it is verified the instability mode is the lower hybrid instability for the investigated cases. 

With the excellent agreement of the real frequency (spectrum) and the growth rate between the theory and the simulations over the investigated magnetic field range, it is confirmed that the instability is of the gradient drift driven instability nature.




\section{Nonlinear saturation and electron heating}

In this section, the short wavelength instability mode transition to the m=1 rotating spoke/macroscopic structure accompanied by an inverse energy cascade is shown. The  electron heating diagram is also addressed.

\subsection{Mode transition accompanied by an inverse energy cascade}

In order to have insight of how the short wavelength mode transits to the m=1 spoke/macroscopic structure, the mode number as a function of time is plotted for magnetic fields $B=20 {\rm mT},40 {\rm mT},60 {\rm mT}$ in Fig. 12. The $\theta - t$ data of the azimuthal electric field $E_{\theta}$ is used and the $\theta$ dimension is transformed to $k_{\theta}$ space by FFT method for plotting the figure. For all three cases, the mode number $m$ decreases with time increasing during the linear-nonlinear transition phase. The $m=1$ rotating macroscopic structure can be formed in the final stage. Particularly when $B=20 {\rm mT}$, in the nonlinear stage, the m=1 macroscopic structure and the shorter wavelength modes coexist, which agrees with the previous study \cite{boeuf2019}.

\subsection{electron heating}

Fig. 13 present $\theta-r$ profiles of potential, mean electron kinetic energy and ionization rate at $B=60{\rm mT}$. From the electron mean energy plot, it is discerned that electrons are heated when they drift in the spoke front double layer and the anode sheath. The electron mean energy $\varepsilon_m$ increases by about $5 {\rm eV}$ along the the spoke front passage, where the $\mathbf{E_{\perp}} \times \mathbf{B}$ electron flow is channeled. Here $E_{\perp}$ is the electric field across the double layer and the anode sheath. The consequence of the heating is the enhanced ionization rate in the downstream region of the spoke front passage. Like the previous study \cite{Boeuf2013}, collisional heating can be important in our cases. According to the classical transport theory, the electron mean velocity parallel to $\mathbf{E_{\perp}}$ and to $\mathbf{E_{\perp}} \times \mathbf{B}$ is:

\begin{equation}
    \mathbf{v_{\perp}}=-\frac{e\mathbf{E_{\perp}}}{m\nu_{en}}\frac{1}{1+h^2}
\end{equation}

\begin{equation}
    \mathbf{v}_{\mathbf{E_{\perp}}\times \mathbf{B}}=\frac{\mathbf{E_{\perp}}\times \mathbf{B}}{B^2}\frac{h^2}{1+h^2}
\end{equation}

\noindent where $h=\omega_{ce}/\nu_{en}$ is the Hall parameter. In the absence of collisions, h becomes infinite, $v_{\perp}$ tends to zero and  $v_{\mathbf{E_{\perp}}\times \mathbf{B}}$ tends to the collisionless limit (electrons are trapped by magnetic field line). With collisions present, electrons drifting parallel to $\mathbf{E_{\perp}} \times \mathbf{B}$ experience collisional transport parallel to $\mathbf{E_{\perp}}$. Electron heating rate is given by $\partial_t \varepsilon_m=-\mathbf{v_{\perp}} \cdot \mathbf{E_{\perp}}$. Assuming the electron $\mathbf{E_{\perp}} \times \mathbf{B}$ drifting time  is t, i.e., drifting over a length of $l=v_{\mathbf{E_{\perp}} \times \mathbf{B}}t$, the electron energy gain is

\begin{equation}
    \varepsilon_m=E_{\perp}\frac{v_{\perp}t}{v_{\mathbf{E_{\perp}}\times \mathbf{B}}t}v_{\mathbf{E_{\perp}}\times \mathbf{B}}t=E_{\perp}l/h
\end{equation}

In our simulations, for $B=60 {\rm mT}$, $E_{\perp}=25 {\rm kV/m}$, $h=42$ and the drifting length (from the anode sheath to the ionization region) $l \approx 7{\rm mm}$ gives $\varepsilon_m=4.2 {\rm eV}$. The energy increase is consistent with the simulation,  showing the important role of the collisional heating. For $B=20 {\rm mT}$, $E_{\perp}=2.5 {\rm kV/m}$, $h=14$ and $l=7 {\rm mm}$ gives $\varepsilon_m=1.25 {\rm eV}$. The small energy increase explains the absence of the enhanced ionization when $B=20 {\rm mT}$ (see Fig. 4).


\section{Conclusion}
In this study, the $m=1$ azimuthally rotating spoke/macroscopic structure in a crossed field plasma was investigated using 2D3V radial-azimuthal PIC/MCC simulations. In the simulations, the formation of a rotating spoke (with enhanced ionization) or macroscopic structure (without enhanced ionization) were clearly observed in the steady state given the instability is excited. The linear-nonlinear transition can be nicely identified, thereby enabling tracing of the source instability which is considered to be the origin/driving mechanism of the rotating large scale structures. 

To analyze the linear development of instabilities in our simulations, the two-fluid dispersion relation of the gradient drift instability was utilized and the space charge effect was introduced in the fluid theory. The super-resolution signal processing method MUSIC was applied to process the simulation data to obtain the frequency spectrum. In the linear stage of the instability, the perturbation spectrum ($\omega-k_{\theta}$) and the growth rate were derived and agree very well with the theoretical dispersion relation of the gradient drift instability across the magnetic field range under investigation. In our cases, the instabilities are induced near the anode where the local radial electric field is positive and aligned with the density gradient, which is the characteristic of the collisionless Simon-Hoh instability. It was further shown that the short wavelength modes are also present in all investigated cases: lower hybrid modes and ion sound modes, which are subject to the space charge effect. Particularly, the most linearly unstable mode was found to be the lower hybrid instability where growth rate is the largest. During the linear-nonlinear transition period, the energy spectrum exhibits cascading from the short to the long wavelengths in all investigated cases. When the spoke/macroscopic structure is formed, the magnitude of $E_{\theta}$ generated in the spoke/macroscopic structure front (double layer) is proportional to the magnetic field strength. The collisional heating of the $\mathbf{E_{\theta}} \times \mathbf{B}$ electrons in the spoke front may explain the formation of the well-established spoke when $B=60 {\rm mT}$, and the absence of the enhanced ionization when $B = 20 {\rm mT}$.

Comparing our model to the other typical cross-field discharges, e.g., magnetron and Hall thruster, several major approximations were made:  reduced plasma density, reduced device geometry and uniform magnetic field. The large plasma density $\sim 10^{19} {\rm m^{-3}}$ in real discharges means extremely small Debye length and the quasi-neutrality condition is therefore expected to hold even in the ion sound instability. The geometry size has most impact on the density gradient length, which is expected to further modify the dispersion relation. The magnetic field gradient renders new free energy to destabilize the gradient drift wave \cite{Frias2012} and enhances the electron heating due to $\triangledown B$ drift \cite{Boeuf2020}. 

\section*{Acknowledgement}
This work has been supported by the German Science Foundation (DFG) within the SFB-TR 87 project framework and by the Research Department ‘Plasmas with
Complex Interactions’ of Ruhr University Bochum. The author L. Xu gratefully acknowledges fruitful discussions with I. Kaganovich and A. Smolyakov. We gratefully acknowledge the suggestions from R. Kleiber on the MUSIC method for obtaining the numerical frequency spectrum.

\appendix
\section{Derivation of the gradient drift instability dispersion relation with the charge separation considered}

The following equations are calculated with Cartesian coordinate and only the mode developed in the azimuthal direction ($\mathbf{y}$) is accounted for as stated in the text. Ions are unmagnetized. The governing equations for cold ions are the mass and momentum conservation equations:

\begin{equation}
     \frac{\partial n_i}{\partial t} +  \triangledown \cdot (n_i \mathbf{v_i}) = 0
\end{equation}

\begin{equation}
     \frac{\partial \mathbf{v_i}}{\partial t} +  (\mathbf{v_i} \cdot \triangledown)  \mathbf{v_i} = -\frac{e}{m_i}\triangledown \phi
\end{equation}

For linear perturbations, $\widetilde{n_i},\phi,\widetilde{v_i} \sim exp[-i(\omega t-k_{\theta}y)]$, the system of Eq. A1 and Eq. A2 reduces to 

\begin{equation}
     \frac{\partial \widetilde{n_i}}{\partial t} + n_0 \frac{\partial \widetilde{v_i}}{\partial y}  = 0
\end{equation}

\begin{equation}
     \frac{\partial \widetilde{v_i}}{\partial t}  = -\frac{e}{m_i}    \frac{\partial \phi}{\partial y}  
\end{equation}

\noindent where $n_0$ is the equilibrium density. Note that in the local approximation, the equilibrium profiles can be considered constant ($\partial_y n_0=\partial_y v_0=0$ where $v_0$ is the equilibrium ion velocity). From A3 and A4, the ion density perturbation $\widetilde{n_i}$ in response to the potential fluctuation $\phi$ can be derived:

\begin{equation}
\frac{\widetilde{n_i}}{n_0}=\frac{k_{\theta}^2}{\omega^2}\frac{e\phi}{m_i},
\end{equation}

The basic equations for electrons are the mass and momentum conservation equations with electron gyro-viscocity, electron inertia and collisions taken into consideration:

\begin{equation}
     \frac{\partial n_e}{\partial t} +  \triangledown \cdot (n_e \mathbf{v_e}) = 0
\end{equation}

\begin{equation}
     n_em_e\frac{\partial \mathbf{v_e}}{\partial t} +  n_em_e(\mathbf{v_e} \cdot \triangledown)  \mathbf{v_e} = en_e(-\triangledown \phi+\mathbf{v_e}\times \mathbf{B})-\triangledown p_e-\triangledown \cdot \Pi - m_en_e\nu_{en}\mathbf{v_e}
\end{equation}

Solution of Eq. A6 and Eq. A7 gives the electron density perturbation \cite{smolyakov2016}:

\begin{equation}
\frac{\widetilde{n_e}}{n_0}=\frac{\omega_d+k_{\theta}^2\rho_e^2(\omega-\omega_E+i\nu_{en})}{\omega-\omega_E+k_{\theta}^2\rho_e^2(\omega-\omega_E+i\nu_{en})}\frac{e\phi}{T_e},
\end{equation}

In the limit of $k_{\theta}\lambda_{De} \gtrsim 1$, the charge separation comes to play a role and requires the solution of the Poisson equation:

\begin{equation}
    \triangledown^2{\phi}=\frac{e}{\varepsilon_0}(\widetilde{n_e}-\widetilde{n_i}),
\end{equation}

Substitute Eq. A5 and Eq. A8 to Eq. A9, we have:

\begin{equation}
\triangledown^2{\phi}=\frac{en_0}{\varepsilon_0}(\frac{\omega_d+k_{\theta}^2\rho_e^2(\omega-\omega_E+i\nu_{en})}{\omega-\omega_E+k_{\theta}^2\rho_e^2(\omega-\omega_E+i\nu_{en})}\frac{e\phi}{T_e}-\frac{k_{\theta}^2}{\omega^2}\frac{e\phi}{m_i}),
\end{equation}

The linearization of the Laplace operator gives the following expression:

\begin{equation}
k_{\theta}^2\phi=\frac{en_0}{\varepsilon_0}( \frac{k_{\theta}^2}{\omega^2}\frac{e\phi}{m_i}- \frac{\omega_d+k_{\theta}^2\rho_e^2(\omega-\omega_E+i\nu_{en})}{\omega-\omega_E+k_{\theta}^2\rho_e^2(\omega-\omega_E+i\nu_{en})}\frac{e\phi}{T_e}),
\end{equation}

After the cancellation and the simplification, the dispersion turns out to be Eq. 2 in the text which reads:

\begin{equation}
(k_{\theta}\lambda_{De})^2=\frac{k^2_{\theta}c^2_s}{\omega^2} - \frac{\omega_{d}+k^2_{\theta}\rho^2_{e}(\omega-\omega_{E}+i\nu_{en})}{\omega-\omega_{E}+k^2_{\theta}\rho^2_{e}(\omega-\omega_{E}+i\nu_{en})}.
\end{equation}

\section*{References}
\bibliography{reference}

\end{document}